\documentclass[sigconf,screen]{acmart}

\usepackage[linesnumbered,ruled,vlined]{algorithm2e}
\usepackage{float}

\usepackage{graphicx}

\usepackage{array}        %
\usepackage{booktabs}     %
\usepackage{tabularx}     %
\usepackage{longtable}    %

\usepackage{enumitem} %

\usepackage[normalem]{ulem}
\useunder{\uline}{\ul}{} 

\copyrightyear{2026}
\acmYear{2026}
\setcopyright{cc}
\setcctype{by}
\acmConference[LAK 2026]{LAK26: 16th International Learning Analytics and Knowledge Conference}{April 27-May 01, 2026}{Bergen, Norway}
\acmBooktitle{LAK26: 16th International Learning Analytics and Knowledge Conference (LAK 2026), April 27-May 01, 2026, Bergen, Norway}
\acmPrice{}
\acmDOI{10.1145/3785022.3785057}
\acmISBN{979-8-4007-2066-6/2026/04}

\begin{document}
 
\title[Toward Trait-Aware Learning Analytics]{Toward Trait-Aware Learning Analytics} 

\author{Conrad Borchers}
\orcid{0000-0003-3437-8979}
\affiliation{
    \institution{Carnegie Mellon University}
    \city{Pittsburgh, PA}
    \country{USA}
}
\email{cborcher@cs.cmu.edu}

\author{Hannah Deininger}
\orcid{0000-0002-8754-6123}
\affiliation{
    \institution{University of Tübingen}
    \city{Tübingen}
    \country{Germany}
}
\email{hannah.deininger@uni-tuebingen.de}

\author{Zachary A. Pardos}
\orcid{0000-0002-6016-7051}
\affiliation{
    \institution{University of California, Berkeley}
    \city{Berkeley, CA}
    \country{USA}
}
\email{pardos@berkeley.edu}

\renewcommand{\shortauthors}{Borchers et al.}

\newcommand{\revision}[1]{\textcolor{black}{#1}}
 
\begin{abstract}
Learning analytics (LA) draws from the learning sciences to interpret learner behavior and inform system design. Yet, past personalization remains largely at the content or performance level (during learner-system interactions), overlooking relatively stable individual differences such as personality (unfolding over long-term learning trajectories such as college degrees). The latter could bring underappreciated benefits to the design, implementation, and impact of LA. In this position paper, we conduct an ad hoc literature review and argue for an expanded framing of LA that centers on learner traits as key to both interpreting and designing close-the-loop experiments in LA. We show that personality traits are \textit{relevant} to LA’s central outcomes (e.g., engagement and achievement) and \textit{conducive} to action, as their established ties to human-computer interaction (HCI) inform how systems time, frame, and personalize support. Drawing inspiration from HCI, where psychometrics inform personalization strategies, we propose that LA can evolve by treating traits not only as predictive features but as design resources and moderators of analytics efficacy. In line with past position papers published at LAK, we present a research agenda grounded in the LA cycle and discuss methodological and ethical challenges. 
\end{abstract}



\maketitle

\section{Introduction}

Learning Analytics (LA) has matured into a socio-technical field that joins computational methods with learning theory to understand and improve learning outcomes \cite{siemens2013learning,teasley2019learning}. Classic definitions emphasized “the measurement, collection, analysis, and reporting of data about learners and their contexts” \cite{conolelak}, but personalization has since become central to how LA is practiced. Recent community updates to LA’s scope explicitly recognize “studies that evaluate the effectiveness and impact of adaptive technologies based on learning analytics” as within scope,\footnote{\url{https://www.solaresearch.org/wp-content/uploads/2025/05/LA-Definitions-for-Community-Feedback.pdf}} accentuating the field’s “close-the-loop” turn toward explanatory and hypothesis-driven accounts.

Yet what counts as \emph{personalization} in LA remains inconsistent. As Pelánek observes, terms such as \emph{adaptive}, \emph{intelligent}, and \emph{personalized} are used heterogeneously across subfields \cite{pelanek2022adaptive}. In practice, most personalization aligns content or feedback to moment-by-moment learner behaviors—error patterns, strategies, help-seeking, or disengagement—detected in system logs \cite{aleven2016instruction,khor2023systematic}. All of these interactions occur within specific time frames of learner-system interactions. What remains underdeveloped is personalization that incorporates \emph{stable, between-learner differences} such as personality traits and other dispositional constructs. These dispositions not only predict learning behavior but also unfold their impact on learning over longer periods of time (e.g., as learners seek out and traverse learning pathways such as completing degrees). Hence, personality constructs are especially relevant to learning contexts such as higher education, where our field increasingly studies long-term learning trajectories (e.g., degree completion, course selection, and course difficulty) that go beyond in-system performance measures (e.g., tutoring system log data). Although some empirical work has created analytics to describe these learning processes (e.g., course workload trends over time \cite{borchers2023insights}), there is a lack of consideration of how and what to personalize related to learner traits.

This gap in consideration of stable learner traits in LA research suggests a missed opportunity: LA’s technical strengths in modeling fine-grained log data could be coupled with explanatory constructs to move beyond descriptive outcomes and toward mechanisms that clarify \emph{why} interventions work, \emph{for whom} they work, and \emph{how} explanatory accounts can guide transferable process models. This integration would not only honor the growing emphasis on theory-driven work in LA but would also connect log-based evidence with psychological constructs that make findings more interpretable and transportable across contexts, thus addressing long-term challenges of LA \cite{baker2019challenges}.

Our field's emphasis on state-like signals as opposed to stable traits matters. A substantial body of research demonstrates that trait-like characteristics reliably predict academic outcomes and learning processes. For example, conscientiousness robustly predicts academic performance beyond cognitive ability \cite{poropat2009meta}, whereas neuroticism and procrastination tendencies are linked to stress responses and state test performance, respectively (e.g., \cite{mammadov2022,gurung2025starting}). Adjacent communities, particularly those in human-computer interaction (HCI), increasingly use psychometrics to inform interface and feedback design \cite{stachl2019opportunities}. In LA, however, who a learner \emph{is} is often inferred implicitly from what a learner \emph{does}. For example, log-based measures of “procrastination” may show trait-like stability across contexts \cite{gurung2025starting} and help explain achievement patterns \cite{borchers2025workload}. However, they capture only a narrow slice of learner differences, as personality traits are most commonly assessed via self-report.

We argue that LA has not fully leveraged the study of between-learner variability. Its emphasis on intraindividual, state-like signals reflects a tradition in the learning sciences of seeking principles that generalize across learners. While such principles are essential, focusing only on the “average learner” constrains external validity, limits transportability, and overlooks traits that could strengthen both explanatory accounts and analytics design decisions.

\subsection{States vs. Traits and Their Roles in LA}

Early research in LA has distinguished state and trait explanations of learner behavior, such as for gaming the system. In particular, Baker \cite{baker2007gaming} compared whether stable learner (trait) effects, as opposed to learning content (state) effects, better explain gaming behavior. Baker found strong evidence for state effects in gaming, and later research has combined both accounts to derive more accurate estimates of gaming the system \cite{huang2023using}. Indeed, for moment-by-moment motivational states, there is evidence that state explanations provide better accounts of learner behavior in tutoring systems \cite{Dang2022}. 

Despite the utility of state definitions to describe in-system learner behavior, not all learning occurs in learning systems bound to a particular time and place. This study takes a broader definition of traits, which encompasses learner attributes that not only manifest within learning software but throughout students' academic careers (e.g., multi-year college degrees). We define \emph{state-like} constructs as dynamic, situation-sensitive variables (e.g., momentary knowledge, affect, disengagement) that enable within-learner, moment-to-moment adaptation. \emph{Trait-like} constructs are relatively stable tendencies (e.g., conscientiousness) that enable reliable between-learner differentiation across contexts and time.

In \emph{Trait-Aware Learning Analytics (TALA)}, traits are not additional predictors but (i) \emph{moderators} that pre-specify \emph{for whom} behaviors and interventions have larger/smaller effects, and (ii) \emph{design primitives} (i.e., basic, reusable building blocks for adaptation that inform pacing, framing, and timing of support). Operationally, TALA models states at the interaction grain and traits at the learner grain (separating within- and between-person variance), and tests preregistered trait $\times$ predictor or trait $\times$ intervention interactions. TALA focuses on constructs with evidence of stability and links to learning. Since survey measures are not always feasible, we discuss validated trace-based proxies, accompanied by reliability and validity checks. Subsequent sections map traits into the LA cycle and illustrate moderation and design implications in three case studies.

\subsection{The Present Study}

We scope our argument to psychological constructs with meta-analytic links to learning outcomes and processes that are commonly measured via questionnaires (e.g., Big Five, goal orientation, self-efficacy). Not all differences are equally stable, useful, or ethical to measure at scale; therefore, we offer specific guidelines and case studies for TALA in common LA settings. We list dispositional constructs robustly associated with learning and scope opportunities for LA in relation to each through a literature review and empirical case studies. As its primary contributions, this position paper:
\begin{itemize}[leftmargin=*]
  \item Articulates a theoretical case for integrating personality and dispositional constructs into LA, clarifying the state–trait distinction and positioning traits as moderators and design primitives.
  \item Proposes the \emph{TALA} framework that embeds traits across the LA cycle to support explanatory, hypothesis-driven experimentation and transportable analytics.
  \item Outlines a research agenda spanning measurement (validated self-report vs. behavioral proxies), analytics (moderation, heterogeneity of treatment effects), and intervention design (trait-aware adaptivity), alongside ethical considerations.
  \item Discusses data-driven insights from case studies combining personality assessment with LA.
\end{itemize}

\section{Related Work}

Before discussing and surveying specific constructs, we situate emerging research related to stable individual differences in LA and its related disciplines to strengthen our claims of novelty. 

\subsection{Current State and Roles of Traits in LA}

Within LA, personalization overwhelmingly targets state-like signals (knowledge, engagement, help-seeking) inferred from logs. At the same time, individual differences have appeared mostly as correlates rather than as drivers of adaptation. A representative example is Matcha et al., who linked Big Five traits to MOOC strategy profiles—valuable descriptively, but not used to route supports or change system behavior \cite{matcha2020}. Recent LAK work has inched closer to stable individual-difference signals via ``timing delay profiles'' (e.g., differentiating subpatterns of deadline-driven delay rather than labeling all delay as procrastination) \cite{kim2025delay}. Yet, these studies explicitly treat delay as a behavioral manifestation shaped by context, not a validated trait instrument suitable for high-stakes personalization. In parallel, work on course planning uses transaction logs and course load analytics (CLA) to quantify workload exposure and show that late enrollment and irregular planning predict risk (e.g., late drops), again emphasizing situational signatures over trait diagnoses and arguing for careful measurement before any trait-sensitive intervention \cite{borchers2023insights,borchers2025workload}.

Some scholars in LA also argue that personality and motivational dispositions significantly influence learning processes and responsiveness to support. The MetaTutor system demonstrates that theoretically grounded individual differences (including motivation and self-regulated learning; SRL) align with how students benefit from scaffolds in intelligent tutoring. It operationalizes those differences to study heterogeneous effects of adaptive support \cite{azevedo2022lessons}. Given that SRL skills are learnable, they fall out of narrower definitions of learner traits. In HCI (an important parent discipline of LA), surveys explicitly argue that personality can be a principled design resource for personalization while highlighting measurement and ethics pitfalls that any deployment should address \cite{stachl2019opportunities}.

Taken together, the LA literature has not yet produced trait-aware, close-the-loop studies that pre-specify validated trait or proxy measures, test moderation prospectively, and adapt timing, framing, or pacing interventions based on those traits, with ethical safeguards. Most prior work uses traits only for post-hoc interpretation \cite{matcha2020}, or models delay and regularity as context-shaped risk signatures rather than trait assessments \cite{kim2025delay,borchers2025workload}. Our contribution is to treat traits as moderators and design primitives across the LA cycle, pairing validated measures with trait-aware adaptivity.

\subsection{Personalization and Traits in Related Fields}

Personalization spans several adjacent communities. In AIED and EDM, most adaptive mechanisms have historically targeted \emph{states}, especially knowledge (and its growth), with some extensions to momentary affect and engagement \cite{aleven2016instruction}. This emphasis on within-learner dynamics yielded impactful models of tutoring decisions, help seeking, and disengagement, but only rarely incorporated relatively stable \emph{traits} such as personality.

There is, however, a growing line of AIED and tutoring system work that brings personality explicitly into the loop. In conversational ITS, researchers have begun to \emph{model and simulate} Big Five profiles and study how these profiles shape tutoring dialogues and scaffolding choices \cite{liu2024personality}. Adjacent poster and late-breaking findings report early deployments of personality-aware conversational tutors in programming courses \cite{alrobai2025personality}. Pedagogical-agent studies—a longstanding strand within AIED—show that agent persona and feedback style interact with learner characteristics, with downstream effects on learning and experience \cite{schroeder2017measuring,kim2016effects,baylor2009nonverbal,johnson2018pedagogical}. Beyond core tutoring behavior, work on \emph{personalized explanations} finds that the usefulness of system explanations is modulated by user characteristics, including cognitive ability, arguing for explanation strategies that adapt to such traits \cite{conati2021toward}.

Motivational dispositions have been explored in AIED, EDM, and LA research. In EDM, studies linked self-reported goal orientation and self-efficacy to in-system behaviors and outcomes, while cautioning that log-based approximations of these traits are limited \cite{fancsali2014goal}. By contrast, some LA work has used log data as proxies for self-efficacy and cognitive load \cite{jovanovic2019introducing}. Yet such dispositions have rarely guided analytics or adaptivity, with a few exceptions proposing difficulty adjustments to sustain motivation \cite{du2016implementation} or content personalization by topic interest \cite{walkington2020appraising}. 

In HCI research, trait-based personalization is well established. Personality has informed recommendation strategies (e.g., tailoring diversity to Openness) and anticipated interaction preferences, trust, and acceptance \cite{tintarev2013adapting,stachl2019opportunities}. In education technology more broadly, Big Five traits have been linked to adoption and patterns of GenAI use \cite{arpaci2025role}, highlighting the practical value of trait-aware design choices even without full learner models.

All of these findings could potentially inform the delivery of instructional interventions, but they are different from TALA (proposed in this article), since TALA argues for \textit{personalizing analytics} and delivery to learner traits (as opposed to instructional decisions). Indeed, LA sits at a productive intersection: it integrates platform traces with institutional and contextual data, evaluates interventions at scale, and considers socio-technical implications. Even when direct personality inventories are absent, LA can recover \emph{quasi-trait} constructs from large-scale process data (e.g., procrastination via delayed starts), providing trait-like signals that are reliable and predictive \cite{gurung2025starting}. Earlier work on behavioral measurement models of motivation likewise shows how stable dispositions can be inferred from behavior over time and then linked to learning processes \cite{Dang2022}. This positions TALA to (i) use survey-based personality measures when available and behavioral proxies when they are not, (ii) study how trait-level differences moderate the effects of analytics-driven feedback and dashboards, and (iii) surface trait-aware supports that inform advising, course selection, and broader learner success infrastructures. We exemplify these applications through case studies below.

\section{Surveying Trait Constructs Relevant to LA}

Learners differ in a range of characteristics referred to as individual differences that can be broadly divided into trait-like and state-like differences. Trait-like differences (e.g., conscientiousness, grit) are relatively stable across time and contexts, enabling interindividual comparisons (i.e., how people differ from each other), which supports forming groups of learners with similar profiles (e.g., \cite{marano2025examining}). In contrast, state-like differences (e.g., current interest, stress, situation-specific motivation) are more dynamic, fluctuating with context and time, and enable intraindividual comparisons (i.e., how the same person varies across situations over time). In the following, we describe a curated list of constructs relevant to LA.

\subsection{Search Methodology}

\revision{Based on a focused and transparent literature search appropriate for an empirical position paper, we curated a theory-informed set of trait-level constructs relevant to LA. Rather than conducting a full systematic review, our approach combined a theory-first scoping with reproducible, targeted searches consistent with conventions in prior LAK position papers \cite{yan2024generative}. First, we scoped psychological traits (not demographics or skills) that plausibly affect learning processes or responsiveness to interventions based on seminal meta-analyses, review papers and overview articles on personality traits by searching for highly-cited articles matching ``personality and academic outcomes'' on Google Scholar \cite{busato2000intellectual,noftle2007personality,jensen2015personality,komarraju2011big,de1996personality}. We then extracted candidate constructs from these studies until no new constructs were introduced by a study.} Second, we ran targeted searches of meta-analyses for each candidate construct (query pattern: \texttt{\{construct\} + meta-analysis + (learning|education|achievement)}). Third, we applied inclusion/exclusion rules: include only constructs with (a) evidence of relative stability, (b) documented links to academic outcomes or core learning processes, and (c) practical implications for LA adaptations; exclude predominantly state-like constructs, domain skills (e.g., SRL strategies), and traits with inconsistent or null links to achievement. Fourth, for each retained construct, we extracted a representative definition, the most conservative effect size(s) from recent syntheses and meta-analyses, and a constructed concrete “analytics implication” (how it could inform features, moderation tests, or adaptations). Table~\ref{tab:overview} reflects the result of this procedure.

\subsection{Overview of Relevant Constructs}
Various categories of trait-like differences, including personality, cognitive abilities, and motivational dispositions, can inform personalization in LA (see Table \ref{tab:overview}). Although well-studied in psychology, these constructs have been mostly assessed using self-report questionnaires, and little research has explored their inference from behavioral trace data \cite{dang2020self}. \textit{Personality traits} are stable characteristics shaping how individuals think, feel, and act. The Big Five framework \cite{digman1990personality}—openness, conscientiousness, extraversion, agreeableness, and neuroticism—is widely accepted. Conscientiousness, marked by organization, dependability, and diligence \cite{roberts2009conscientiousness}, is consistently associated with positive outcomes, including academic achievement \cite{mammadov2022}, job performance \cite{barrick1991big}, and well-being \cite{anglim2020predicting}.

While traditionally studied through self-report questionnaires, personality traits are increasingly being considered in LA research. For instance, \cite{matcha2020} examined how learning strategies derived from behavioral sequence data in a MOOC setting relate to the Big Five personality traits. Other studies have focused on inferring personality traits directly from behavioral trace data. For example, \cite{raemy_measuring_2025} investigated whether behavioral indicators of learning could predict conscientiousness-related traits and academic performance. Their findings showed that behavior-based indicators were moderately correlated with self-reported conscientiousness and also predicted academic performance, supporting the potential of using personality-related constructs to enhance learner modeling and personalization in digital learning environments. However, research explicitly integrating personality traits into LA applications to enhance learning processes or interventions remains relatively scarce.

Motivation is often defined as the internal drive that determines how much effort an individual is willing to invest in pursuing a goal. While motivation can fluctuate across contexts and tasks, some dispositions, such as goal orientation, are considered relatively stable over time and can thus be treated as trait-like. These dispositions influence how students approach learning tasks, the strategies they employ, and how they respond to instructional interventions \cite{koeller2012, schneider2017}. 

Motivational constructs have drawn growing attention, with goal orientation as a central example. It distinguishes mastery-oriented learners, who aim to develop competence, from performance-oriented learners, who seek to demonstrate their competence relative to others, and has been shown to predict outcomes across diverse contexts \cite{koccak2021factors, wirthwein2013achievement}. For example, Choi et al. found notable misalignments between learners’ self-reported achievement goals and their logged behaviors, suggesting that stated goals may not reliably reflect enacted strategies \cite{choi2023logs}. In LA research, however, motivation is typically treated as dynamic and context-sensitive rather than a fixed disposition. As a result, most studies emphasize state-level processes, such as temporal fluctuations in motivation or the role of motivational regulation within broader SRL frameworks \cite{azevedo2022lessons}.

Cognitive abilities refer to the mental capacities required for learning, including reasoning, memory, attention, and problem-solving skills. Individuals differ in these abilities—for example, in their general intelligence, which reflects their capacity to reason, learn, and adapt, or in the amount of cognitive load they can effectively process. While intelligence is typically considered a stable trait, LA research has primarily focused on state-level approximations of cognitive ability, such as momentary cognitive load \cite{larmuseau2020multimodal} or effort \cite{nkomo2021}, which can be inferred from behavioral trace or physiological data.

However, integrating trait-level cognitive abilities into LA systems is equally important. Learners' general cognitive capacity can influence how they interact with digital environments, how much support they need, and how quickly they progress through material. Considering such stable characteristics could inform long-term personalization strategies, such as task sequencing, pacing, or the level of scaffolding provided.

\begin{table*}[t]
\footnotesize
\renewcommand{\arraystretch}{0.94}
\caption{Individual Differences Relevant to LA. \emph{S/C} = Stability / Context-specificity (Y = yes, N = no, — = not specified).}
\label{tab:overview}

\begin{tabularx}{\textwidth}{p{1cm} p{1.4cm} p{2.3cm} X p{0.6cm} p{1.1cm} p{1.0cm}}
\toprule
\textbf{Category} & \textbf{Construct} & \textbf{Definition} & \textbf{Relevance to Learning and LA} & \textbf{S/C} & \textbf{Effect} & \textbf{Citation} \\
\midrule

Personality & Conscientious-ness &
Organized, responsible. &
Better planning, time management, consistent study habits, and self-discipline. → adjust pacing and deadlines. &
Y/N & 0.56 / 0.54 &
\cite{mammadov2022,koccak2021factors,richardson2012psychological,trapmann2007meta,vedel2014big} \\

 & Openness &
Open to new aesthetic, cultural, and intellectual experiences. &
Preference for deep, creative approaches. → Offer exploratory tasks, diverse examples, and enrichment. &
Y/N & 0.16 / 0.28 &
\cite{mammadov2022,koccak2021factors,richardson2012psychological,vedel2014big} \\

 & Agreeableness &
Cooperative, prosocial. &
Supports teamwork and collaboration. → Match to cooperative groups; collaborative tasks. &
Y/N & 0.09 &
\cite{mammadov2022,vedel2014big} \\

 & Neuroticism &
Emotional instability. &
Higher stress; disengagement risk under pressure. → Provide stress-mitigation prompts.&
Y/N & -0.02 &
\cite{mammadov2022} \\

 & Grit &
Perseverance toward long-term goals. &
Sustained effort despite setbacks. → Progress tracking and rewards. &
—/— & 0.29 &
\cite{koccak2021factors,crede2017much} \\

 & Self-concept &
Self-evaluation of abilities. &
Higher confidence leads to greater persistence. → Adapt task difficulty framing; flag mismatches between self-concept and actual performance. &
—/Y & 0.23 &
\cite{koccak2021factors,ma1997assessing} \\

 & Curiosity &
Drive to investigate novel, interesting material. &
Increases attention and depth of engagement. → Trigger curiosity-stimulating prompts; include open-ended or exploratory resources for exploration. &
—/Y & 0.30 (intellectual investment) &
\cite{vonstumm2011hungry} \\

 & Need for Cognition &
Enjoyment of effortful thinking. &
Preference for deep thinking and problem-solving. → Provide challenging, thought-provoking tasks. &
—/Y & 0.39 &
\cite{koccak2021factors,richardson2012psychological} \\

 & Critical thinking &
Evaluative, problem-focused thinking. &
Supports quality of reasoning and solutions. → Scaffold by giving room for argumentation and reflection. &
—/— & 0.50 &
\cite{koccak2021factors,richardson2012psychological,fong2017meta} \\

 & Locus of control &
Perceived control over outcomes (internal vs. external). &
Links to motivation. → Frame feedback differently depending on perceived locus of control. &
—/Y & 0.26 &
\cite{koccak2021factors,richardson2012psychological} \\

 & Procrastination &
Delaying intended actions. &
Last-minute work; reduced performance. → Reminders; chunking; progress gamification. &
—/— & -0.36 &
\cite{koccak2021factors,richardson2012psychological,kim2015relationship} \\
\midrule

Motivation & Goal orientation &
Mastery vs.\ performance approach/avoid. &
Focus of effort and strategy use. → Tailor feedback: process- vs.\ outcome-focused.  &
—/Y & 0.23 / 0.13 (mastery); 0.06 (perf.) &
\cite{koccak2021factors,wirthwein2013achievement} \\

 & Self-efficacy &
Belief in the capability to succeed. &
Higher confidence leads to persistence and effort. → Adaptive encouragement; scaffolded success. &
—/Y & 1.17 (0.74 academic) &
\cite{koccak2021factors,sarier2016factors} \\

 & Intrinsic motivation &
Interest/enjoyment of the activity itself. &
Sustained engagement and depth. → Meaningful tasks; autonomy support; gamified goals. &
—/Y & 0.31 / 0.26 &
\cite{koccak2021factors,richardson2012psychological} \\

 & Interest &
Stable preference for activities. &
Attention and deep engagement. → Personalized examples; adaptive recommendations based on interests. &
—/Y & 0.31 &
\cite{schiefele1992interest} \\

 & Academic goals &
Target proficiency within time window. &
Directs attention and effort. → Align tasks and trajectories with academic goals. &
—/Y & 0.31 &
\cite{koccak2021factors,robbins2004psychosocial} \\
\midrule

Cognitive Abilities & Intelligence (Gf) &
Reasoning, learning, adaptation. &
Supports strategy use and efficiency.  → Adjust cognitive load of tasks; differentiate instruction pace. &
N/— & 0.38–0.41 &
\cite{peng2019meta} \\\bottomrule
\end{tabularx}
\end{table*}

\subsection{Data Collection Considerations}

In trait-aware LA, credibility hinges on measurement quality. Survey-based assessments (e.g., BFI-2 and its short forms) provide constructs with documented internal structure and reliability, whereas ultra-brief scales prioritize efficiency and yield lower psychometric precision \cite{soto2017next,GoslingTIPI}. Treating a variable as trait-like presupposes evidence of rank-order stability at the study’s time scale, typically demonstrated via test–retest or intraclass correlations \cite{roberts2000rank}. Cross-lingual or cross-cultural applications further require evidence for measurement invariance to support comparisons or profile-based analyses \cite{schmitt2007geographic}. When surveys are infeasible, longitudinal trace features—such as multi-term delay or regularity indices—can function as proxies, but only insofar as convergent validity against gold-standard surveys and temporal stability are established on a labeled subsample, with uncertainty reported explicitly.

Analytically, separating within- from between-learner variance (e.g., appropriate random-effects structures) reduces state contamination of trait estimates, while reliability-aware estimation (e.g., attenuation corrections) contextualizes moderation magnitudes. Nonresponse and selection into trait availability shape generalizability and should be summarized alongside the main results. Because behavioral proxies can reflect contextual constraints (e.g., work schedules affecting “delay”), fairness considerations pertain to both measurement error and downstream impact. Across these choices, a purpose-limited, opt-in procedure with transparent retention and revocation policies aligns trait use with prevailing ethics guidance and regulatory expectations, since personality data are considered sensitive. 

Another key consideration is to check for algorithmic bias when stratifying interventions based on stable traits, especially when those traits may correlate with demographic characteristics. For instance, all Big Five traits (except openness) have been shown to differ by gender on average based on large-scale and cross-cultural surveys \cite{schmitt2008can}, raising the risk that a personalization strategy keyed to these traits could inadvertently replicate gender disparities in access to support. Similarly, using behavioral proxies for procrastination may confound structural constraints (e.g., work schedules that affect first-generation or low-income students) with stable dispositions. LA offers established methods for studying the disparate impact of interventions and algorithmic fairness \cite{mangal2024implementing}. We will return to this issue in the discussion.

Personalizing by relatively stable traits (e.g., neuroticism) risks \emph{indirectly} linking support decisions to protected characteristics, since some trait measures or behavioral proxies differ across demographic groups or reflect contextual constraints. To mitigate these risks, data should be collected only on an opt-in basis with a clear purpose and minimal scope. Researchers should also test for measurement invariance or differential item functioning to ensure trait measures function equivalently across groups. Decisions should not rely on rigid thresholds tied to a single trait, and evaluations of trait-based support should report whether benefits are distributed equitably, alongside calibration within groups \cite{mangal2024implementing}. For instance, suppose we use a stable ``procrastination'' trait, often approximated by multi-term enrollment delay \cite{kim2025delay}, to determine which students receive early pacing nudges. If students who work off campus, who are disproportionately first-generation and from minority groups, tend to register later because of job-related constraints, this proxy may flag them more frequently even when their underlying learning needs are similar to those of other students. Without appropriate safeguards, this could contribute to inequitable intervention outcomes and access. A safer design would (a) validate that enrollment delay truly reflects planning behavior rather than work constraints using a labeled subsample, (b) provide clear opt-out mechanisms, and (c) audit whether nudge accuracy and outcomes are comparable across demographic groups.

\section{Toward Traits as Part of the Learning Analytics Cycle}

Following prior LAK position papers \cite{yan2024generative}, we frame our argument through the LA cycle \cite{clow2012learning}—learner, data, analytics, and intervention—and outline a vision for TALA. We contend that traits are both \textit{relevant} to LA’s goals and systems and \textit{conducive} to design, given their theoretical grounding, measurability, and ability to predict learning.

\subsection{Learner: Specify \emph{Who} and \emph{Why}}
To integrate traits into LA, we first specify \emph{which} individual differences matter for the target task and \emph{why}. Personality traits (e.g., Big Five) are among the strongest non-cognitive predictors of academic performance, especially conscientiousness, which predicts achievement beyond cognitive ability \cite{poropat2009meta,mammadov2022,richardson2012psychological,trapmann2007meta,vedel2014big}. Other constructs differ in stability and domain-specificity (e.g., self-efficacy), but all can shape how learners engage with content, respond to feedback, and benefit from interventions. In TALA, traits are treated as \emph{moderators} (explaining for whom analytics/interventions work) and as \emph{design resources} (informing pacing, framing, and choice architecture), rather than as mere predictive features as is often the case in past research. Design decisions at the Learner stage include:
\begin{itemize}[leftmargin=*]
\item \textbf{Select traits by theory and use-case.} From Table~\ref{tab:overview}, choose a minimal set with clear theoretical links to the focal outcome or process (e.g., conscientiousness for time management; goal orientation for feedback uptake).
  \item \textbf{State vs. trait.} Pre-specify which constructs will be modeled as trait-like (between-learner) versus state-like (within-learner).
  \item \textbf{Hypothesize heterogeneity.} Articulate moderation hypotheses (e.g., ``the benefit of workload smoothing is larger for low conscientiousness'') to guide modeling and intervention design.
\end{itemize}

\subsection{Data: Elicit and Validate Trait Signals}

Unlike clickstream behaviors, traits are commonly measured via validated self-report instruments; they can also be approximated with log data (e.g., multi-term delay indices) when surveys are impractical. TALA adopts a \emph{dual-sourcing} strategy (i.e., surveys where feasible, trace-based proxies where needed) paired with validation based on survey measures in smaller samples. The stronger a proxy’s validity, the more likely it is to adequately represent a construct of interest and can be used for assessing student-level personality. Lower-fidelity proxies can still be used for analytics of institutional trends, for instance, for curriculum analytics and redesign. Design decisions at the Data stage include:
\begin{itemize}[leftmargin=*]
  \item \textbf{Measurement plan.} Specify (i) survey instruments and (ii) proxy features (e.g., across-semester stability of enrollment delay) with a rationale grounded in the construct's theory.
  \item \textbf{Stability and validity.} Test longitudinal stability (e.g., Cronbach's $\alpha$, ICC) and, for behavioral proxies, convergent validity against surveys on a subsample before operational use.
  \item \textbf{Temporal decomposition.} Separate within- from between-learner variance (e.g., person-mean centering).
\end{itemize}

\subsection{Analytics: Model Heterogeneity}
TALA centers its analytics on heterogeneity: traits help explain variation in outcomes and responsiveness to interventions. Models capture moderation explicitly and, ideally, test whether discovered patterns generalize. Design decisions at the Analytics stage include:
\begin{itemize}[leftmargin=*]
  \item \textbf{Confirmatory moderation.} Preregister trait $\times$ predictor (or trait $\times$ intervention) interactions in multilevel models, e.g.,
  \[
  y_{ict}=\beta_0+\beta_1\,X_{ict}+\beta_2\,\textsc{Trait}_i+\beta_3\,X_{ict}\!\times\!\textsc{Trait}_i + u_i + v_c + w_t + \varepsilon_{ict},
  \]
  with random effects for learner ($u_i$), context ($v_c$), and measurement time ($w_t$).
  \item \textbf{Heterogeneous treatment effects.} Where randomized or quasi-experimental interventions exist, estimate CATE/uplift; audit discovered learner segments (e.g., prior knowledge differences).
  \item \textbf{Transportability checks.} Optionally, validate moderation patterns across contexts (e.g., courses, cohorts), and document where trait–effect relations hold or shift.
\end{itemize}

\subsection{Intervention: Close the Loop Using Traits}
Analytics and models inform adaptivity, which is prospectively evaluated on new learners to close the loop. TALA focuses on \emph{when} to intervene, \emph{how} to frame support, and \emph{how much} to adjust, all conditioned on trait profiles or validated proxies. Design decisions at the Intervention stage include:
\begin{itemize}[leftmargin=*]
  \item \textbf{Adaptation surfaces.} 
  \emph{When to intervene} (e.g., high workload in combination with procrastination signals).
  \emph{How to intervene} (e.g., process vs. performance emphasis by goal orientation).
  \item \textbf{Explainability \& agency.} Providing profile-contingent rationales (``given your historical planning pattern, starting two weeks earlier reduces drop risk by $x\%$'') and user controls (e.g., opt-out).
  \item \textbf{Evaluation and equity.} Run experimental trials stratified by trait segments; report average and segment-specific impacts, plus \emph{parity of benefit} across demographic groups.
\end{itemize}

\section{Case Studies}

We present three case studies spanning advising, intelligent tutoring, and course planning to show how incorporating trait-level constructs can reveal heterogeneity in behavior–outcome relations and motivate actionable, ethically scoped adaptations.

\subsection{Case Study \#1: Trait-Aware Advising}

The first case study centers around using Big Five traits to determine how course workload analytics could be used for academic advising (i.e., encouraging vs. discouraging high loads).

\paragraph{Setting and Approach}

This case study synthesizes a trait-aware advising pattern from research on course planning \cite{BorchersPardosPerformanceWorkload}. The underlying studies integrate (i) registrar enrollment transactions with course-level \emph{workload} estimates (time load, mental effort, psychological stress) derived from learning management system and enrollment features, (ii) brief, validated \emph{trait} measures (e.g., NEO-FFI conscientiousness, neuroticism; academic self-efficacy short forms) collected under opt-in consent \cite{BorchersPardosPerformanceWorkload}, and (iii) advising \emph{outcomes} such as normalized term grades. The analytic focus is \emph{moderation}: the question is not whether heavier loads lead to lower grades on average, but \emph{for whom} they do. Operationally, enrollment logs are exported each term from the student information system; workload predictions are produced using previously validated models replicated from prior research \cite{borchers2023insights}. Trait and self-efficacy short forms are administered online and linked via privacy-preserving identifiers. Analyses rely on preregistered linear mixed-effects models with student random intercepts and trait $\times$ workload interactions.

\paragraph{Results}
Personality added a moderation effect: A three-way interaction (conscientiousness×neuroticism×workload) was significant, indicating that the workload–achievement benefit generally held except when both conscientiousness and neuroticism were comparatively low. Notably, self-efficacy was positively associated with choosing higher semester workloads and achieving higher grades.

\paragraph{Implications for TALA}
In TALA-informed advising, workload forecasts can be paired with profile-contingent supports: learners with jointly low conscientiousness and low neuroticism tend to benefit from conservative loads with added structure, while profiles with high conscientiousness \textit{or} neuroticism often tolerate heavier loads. Beyond the load itself, the presentation of analytics may matter: for learners higher in neuroticism, sustained achievement under high workload may partly reflect stress-driven effort, so CLA outputs could be contextualized to emphasize supportive pacing, recovery windows, and manageable milestones rather than urgency or deadline cues. Degree-planning dashboards can therefore present explainable workload predictions for the proposed schedule alongside a learner-controlled fit indicator and, where workload manageability is lower, surface scaffolds such as workload planning tools. This matters because advisors often uniformly discourage course overloads \cite{mckinney2024advise}.

\subsection{Case Study \#2: Motivation and Achievement}

\paragraph{Setting and Data}
This study draws on a cluster-randomized field trial of foreign language learning with an intelligent tutoring system (ITS) in German secondary schools \cite{deininger2025whodidwhat}. A total of 618 students from 24 seventh-grade classes across seven academic-track schools participated in 2021/2022. Excluding non-users yielded a final sample of 507 students (55.8\% female, $M_{age}$ = 12.5, $SD_{age}$ = 0.41). The ITS complemented English instruction with structured grammar and vocabulary practice across four task cycles; this analysis focused on cycle 1 to reduce attrition effects.

Students completed a prior knowledge test before working with the system. Additionally, self-report questionnaires assessed prior English knowledge, motivational constructs based on expectancy-value theory \cite{eccles2023expectancy}, and personality traits (openness and conscientiousness). During the task cycle, behavioral trace data from $\sim$222,000 student–system interaction events were collected and engineered into theory-informed learning indicators, capturing frequency, duration, and distribution of practice. After the task cycle, students completed an English proficiency posttest.

\paragraph{Modeling the Moderational Effect of Motivation on the Learning Behavior-Achievement Relationship}
To examine whether motivation moderates the link between learning behavior and achievement, we trained machine learning models to predict posttest English proficiency from behavioral indicators. The best model (XGBoost) explained $R^2 = .41$ of the variance. To interpret predictions, we computed SHAP values indicating each feature’s contribution to outcomes and clustered students by similarity in these values. This produced three learner groups that differed in behavior–achievement relationships, posttest performance (high, medium, low), motivational profiles (self-concept, utility value, perceived cost), and prior knowledge. Notably, students with lower prior knowledge but high versus low motivation engaged with the system differently and achieved correspondingly better or worse outcomes, showing that identical behaviors could predict opposite outcomes depending on their dispositions. These findings suggest that motivation shaped system use and, together with prior knowledge, moderated how behaviors translated into achievement, highlighting the importance of individual dispositions in LA models of academic performance.

\paragraph{Aligning Interventions With Individual Differences}
The results indicate that theory-grounded individual differences (e.g., motivation, prior knowledge) shape not only achievement but also learning behaviors. We identified three learner profiles with distinct behavior–achievement relations despite often similar study actions. For instance, high-knowledge students in Cluster~1 achieved less when they frequently corrected answers after feedback, suggesting errors driven by haste rather than knowledge gaps. For these learners, metacognitive prompts that nudge more careful first attempts (e.g., ``Take a moment to double-check before submitting'') may be more effective than additional content support.

By contrast, students in Cluster 2 entered with lower prior knowledge but moderate motivation, and their performance improved when they received frequent feedback and corrected mistakes. For them, corrections signal productive learning rather than struggle. These students appear well served by the system’s existing feedback and require little intervention.

In Cluster 3, however, corrections after feedback were negatively associated with achievement, despite similarly low prior knowledge as in Cluster 2. These students also reported lower motivation, and behavioral data revealed reduced time-on-task and increased time spent on non-instructional pages. This suggests a pattern of superficial engagement and highlights a need for motivational or engagement-enhancing interventions—such as game-based elements or learning-to-learn supports aimed at boosting strategic engagement \cite{bernacki2020b}.

From a theoretical perspective, these findings support the idea that motivational dispositions moderate the effectiveness of behavioral strategies, a claim often proposed in educational psychology (e.g., \cite{bakhtiarvand2011moderating}) but rarely tested using behavioral trace data. From a practical standpoint, the results show that behavioral interventions should not be based solely on average patterns across all students. Instead, effective personalization requires interpreting student-system interactions in the context of learners' individual differences, such as motivation and prior knowledge. These findings underscore the potential of using trait-level constructs to determine which learners benefit from which forms of feedback, paving the way for more targeted, adaptive, and theoretically grounded interventions.

\subsection{Case Study \#3: Course Planning}

\paragraph{Setting and Data}
This study analyzes course “shopping” using fine-grained enrollment \emph{transactions}—adds, drops, waits, and swaps—across institutional phases (Phase~1, Phase~2, add/drop, late add/drop) \cite{borchers2025workload}. These traces are coupled with CLA, which estimates time load, mental effort, and psychological stress from LMS and enrollment features and aggregates them to term-level workload \citep{borchers2023insights}. Planning behavior is summarized by two longitudinal indices: a \emph{delay} index (greater values indicate actions occurring closer to deadlines) and a \emph{regularity} index (greater values indicate more even spacing). Outcomes include \emph{late drops} (withdrawals after the institutional deadline).

\paragraph{Findings}
Analyses indicate that later \emph{enrollment} timing is associated with \emph{more} late-dropped units and higher \emph{regularity} in academic planning is associated with \emph{fewer} late drops. Workload-focused analyses show that greater delay co-occurs with carrying \emph{more} courses and \emph{higher} total workload at the add/drop deadline; when late drops occur, students disproportionately withdraw from higher-workload courses, consistent with CLA validity. Cross-lagged models support directionality: higher delay in term~$t$ predicts more late-dropped units in term~$t{+}1$, whereas late drops in term~$t$ do not predict subsequent delay in enrollment decisions the next semester.

\paragraph{Interpreting Delay}
In the procrastination literature, trait procrastination is a stable disposition measured with validated questionnaires, whereas enrollment \emph{delay} is a large-scale behavioral correlate—useful for description and personalization but not itself a trait; a single-term delay may also reflect situational constraints such as work schedules, advising holds, or housing \citep{steel2007nature,kim2015relationship}. A trait-aware interpretation instead requires multi-term timing patterns (stability, regularity) linked to \emph{observed} outcomes. Combined with CLA, delay and regularity indices help distinguish \emph{when} workload accumulates and \emph{which} semester course baskets face higher late-drop risk. Academic advisors typically see only observable drops, not planning regularity. Trait-aware analytics could therefore summarize multi-term timing and workload patterns, estimate late-drop risk conditional on CLA and timing profiles, and provide interpretable, outcome-linked indicators that complement existing analytics for advising (e.g., credit hours).

\section{Discussion and Outlook}

This position paper has advanced the case for Trait-Aware LA (TALA), an agenda that foregrounds relatively stable learner dispositions (e.g., conscientiousness, neuroticism, goal orientation) not merely as predictive features, but as moderators and design primitives in the LA cycle. We now reflect on how this proposal connects with prior work, the contributions it makes to theory and practice, and the open challenges that should be addressed.

\subsection{\revision{Ethical Considerations}}

\revision{Incorporating trait-level constructs into learning analytics raises a set of ethical considerations. Many of these concerns are shared with learner modeling and large-scale educational data use, including privacy, informed consent, and proportionality \cite{drachsler2016privacy}. Trait-aware approaches, however, render these issues more salient by foregrounding relatively stable characteristics of learners. Because individuals cannot readily alter their personality traits, interventions grounded in trait information intersect with longstanding questions in the algorithmic fairness literature regarding the distribution of benefits and harms across groups defined by attributes that are not freely chosen \cite{friedman1996bias}. These challenges call for both the application of established methods for identifying and mitigating algorithmic bias \cite{verger2023your} and careful reflection on how trait-based personalization, if designed appropriately, might reduce heterogeneous treatment effects and improve effectiveness across a broader range of learners rather than amplifying existing disparities.}

Trait-aware learning analytics also raises privacy concerns. Personality inventories and trait-anchored behavioral proxies, such as multi-term signatures of delay, constitute personal data that may be sensitive and susceptible to misuse. Ethical deployment, therefore, requires a purpose-limited and opt-in approach that restricts data collection to what is strictly necessary, offers learners meaningful access, correction, and deletion rights, and clearly communicates governance policies regarding collection, retention, and revocation. Such practices align with established guidance in LA ethics and relevant regulatory frameworks \cite{slade2013learning}.

Measurement decisions themselves carry ethical implications. Self-report instruments may be intentionally distorted under incentive, while ultra-short scales often sacrifice reliability for efficiency. Researchers should therefore plan for reliability and validity checks and rely on well-validated short forms when survey burden is a constraint \cite{ViswesvaranOnes1999,Gnambs2014}. Behavioral proxies inferred from trace data warrant particular caution: when used to support confirmatory claims or to drive interventions, they should be validated against appropriate gold standards. Moreover, because trait measures and their proxies can be correlated with protected characteristics through contextual factors, such as employment obligations influencing observed delay, practitioners should examine differential error and differential benefit across groups, report patterns of nonresponse, and avoid applications that are stigmatizing or unnecessarily restrictive, such as using traits to gate access to courses or programs.

Algorithmic fairness presents a further ethical concern for trait-aware learning analytics. Stratifying interventions by relatively stable traits risks reproducing demographic disparities when those traits, or their behavioral indicators, are unevenly distributed across gender or other protected groups \cite{schmitt2008can}. Addressing this risk requires fairness audits that go beyond overall accuracy to examine calibration, subgroup validity, and parity of benefit \cite{mangal2024implementing}. Interventions should be framed as supportive rather than prescriptive, with transparency about why particular recommendations are made and with genuine opportunities for learners to decline or disengage.

Questions of resource allocation, such as how limited advising or instructional support is ultimately distributed in high-stakes settings, fall outside the scope of this position paper. Although trait-aware learning analytics may influence how risk or need is estimated, allocation decisions are fundamentally normative and institutionally situated. Our emphasis is instead on ensuring that the analytics informing such decisions are interpretable, appropriately bounded, and attentive to learner heterogeneity, thereby supporting responsible human judgment. In this sense, insufficient personalization can be as ethically consequential as excessive or poorly designed personalization.

\subsection{Validity Considerations}

Looking ahead, a central set of open problems concerns how construct validity for trait measures and their behavioral proxies can be established and sustained under authentic institutional conditions. A first challenge is that proxy–construct mappings remain insufficiently specified. While behavioral footprints often correlate with psychological traits \cite{youyou2015computer}, the field lacks shared protocols for local validation, such as the use of labeled subsamples, as well as principled approaches for assessing whether these mappings transport across cohorts, courses, or institutions \cite{borchers2023insights}. A second concern is validity under distributional shift. Measurement properties may drift as platforms evolve, policies change, or student populations turn over, yet such drift is rarely quantified \cite{borchers2025generalizability}. Third, cross-context comparability demands more than face validity. Without systematic tests of measurement invariance, particularly for short-form instruments, differences in trait levels or decisions to route learners based on traits risk reflecting artifacts of measurement rather than substantively meaningful variation \cite{schmitt2007geographic}. Finally, differential validity and fairness cannot be assumed. Proxy accuracy, reliability, and missingness may vary systematically across subgroups, making empirical evaluation a prerequisite for responsible use. Progress on these fronts is necessary to accumulate into reproducible and cumulative evidence for trait-aware learning analytics.

A closely related validity issue concerns trait stability and the practical burden of reassessment. Although many dispositional measures exhibit substantial rank-order stability, this stability is weaker in younger populations and can change across developmental periods \cite{roberts2000rank}, implying that more frequent measurement may be warranted when studies span multiple school years. Meta-analytic evidence indicates that stability increases through adolescence before plateauing in adulthood \cite{bleidorn2022personality}. In K-12 settings in particular, lower stability strengthens the case for periodic reassessment and for transparent reporting of test-retest reliability when traits are used to inform personalization or interpretation.

\revision{At the same time, repeated survey administration at scale raises feasibility and ethical concerns. Excessive testing can induce respondent fatigue, degrade data quality, and compete with instructional time. One common response is to rely on well-validated short forms, yet brevity comes at the cost of psychometric precision. Single-item and ultra-short measures may be inappropriate for routing interventions unless their reliability and validity, including test-retest properties, are demonstrated in the local context \cite{soto2017next,GoslingTIPI}. An alternative and potentially complementary strategy is to use the inferential capacity of LA to reduce measurement burden by integrating surveys with trace-based estimates that are explicitly treated as proxies and validated against labeled subsamples. For example, survey-based measures of procrastination can be paired with longitudinal indicators of timing behavior available at scale, such as delay and regularity across multiple terms, to establish convergent validity and to quantify uncertainty before such signals are used operationally \cite{borchers2025workload}. More broadly, combining trait measures with behavioral traces through machine learning may enhance predictive performance in LA models. Recent work suggests that non-linear models can capture additional variance in academic achievement associated with facets of conscientiousness \cite{chernikova2025improving}. Taken together, these considerations point to a pragmatic principle for TALA: traits should be measured as often as required to preserve validity, but no more than necessary to limit burden, with behavioral proxies and machine learning used cautiously to support scalable and responsible intervention.}

\subsection{Other Future Research Directions}

LA is particularly well-positioned to advance trait-aware inquiry because it routinely integrates fine-grained behavioral traces with contextual and institutional data at scale. This integration creates opportunities for closed-loop experimental designs in which learner traits are modeled explicitly as prospective moderators of intervention effects, rather than as post hoc correlates. Such a moderation-first orientation extends early calls for explanatory modeling in LA \cite{clow2012learning} and resonates with recent work in intelligent tutoring systems demonstrating that personality and related traits shape learners’ uptake of feedback and adaptive support \cite{conati2021toward,azevedo2022lessons}. In this way, LA can function as a connective layer between educational psychology’s theoretical accounts of individual differences and HCI’s emerging practices of trait-based personalization \cite{stachl2019opportunities}.

\revision{A second set of opportunities arises from the substantial evidence that many trait–outcome associations replicate across contexts and samples \cite{mammadov2022,koccak2021factors,richardson2012psychological,trapmann2007meta,vedel2014big}. From this perspective, TALA offers a complementary pathway for addressing longstanding concerns about model generalizability and transferability in LA \cite{baker2019challenges}. When predictive models trained in one population are deployed in another, even modest measurements of trait distributions in both settings could provide leverage for diagnosing and adjusting for population shift. Incorporating traits as covariates or moderators, or using them to calibrate predictions to differences in learner composition, may improve transportability without requiring extensive new data collection. At the same time, traits supply a psychologically grounded vocabulary for explaining why the same behavioral features can acquire different meanings across contexts. Prior work, for example, has shown that perceptions of course workload and the predictive utility of learning management system and enrollment features to infer workload vary across institutions \cite{borchers2025generalizability}. Under a TALA framework, such variation can be examined as partially arising from differences in population trait composition or trait-by-context interactions, rather than being treated solely as idiosyncratic properties of individual datasets. More broadly, modeling traits as relatively stable moderators offers a principled way to distinguish behavioral regularities that generalize across settings from those that are contingent on learner composition, supporting LA models that not only transfer more reliably but also yield explanations that are interpretable beyond a single context.} 

\section{Conclusion}

This study proposed \emph{TALA} as a research direction for the field of LA. TALA positions relatively stable learner dispositions as explanatory moderators of behavior–outcome relations, the effectiveness of interventions, and as design primitives for analytics support across the LA cycle. A secondary contribution of this work is an overview of trait constructs relevant for learning, coupled with prospective applications to LA, grounded in an ad hoc literature review. We outlined a dual-sourcing measurement strategy that pairs validated self-reports with trace-anchored proxies. Our case studies in advising, course planning, and intelligent tutoring showed how conscientiousness, neuroticism, motivation, and prior knowledge can reveal heterogeneity that average effects conceal, making interventions more interpretable and actionable for LA researchers and practitioners. Separating within- from between-learner variance clarified \emph{why} the same behavior (e.g., corrections after feedback or high college workload) helps some learners and harms others; reliability-aware estimation and transportability checks can indicate \emph{where} such relations hold across cohorts and contexts; and minimal, opt-in adaptations can demonstrate \emph{how} trait-informed analytics can preserve agency while improving support.

Looking ahead, the path to cumulative science runs through rigorous measurement and responsible deployment. Using TALA, close-the-loop evaluations should preregister hypotheses specifying how intervention effects vary as a function of learner traits, report average and segment-specific impacts, and assess parity of benefit across groups under purpose-limited, opt-in governance. If the community adopts this stance, TALA can move LA beyond prediction toward mechanism-based and principled intervention, yielding supports that extend beyond learner-system interactions to broader and longer-term academic pathways, throughout which trait-based differences in learning outcomes manifest.

\bibliographystyle{ACM-Reference-Format}
\bibliography{main}

\end{document}